\documentstyle[12pt,aaspp4]{article}

\begin{document}

\title{CCD PHOTOMETRY OF GALACTIC GLOBULAR CLUSTERS V. NGC 2808}
\author{Alistair R. Walker}

\affil{Cerro Tololo Inter-American Observatory, National Optical Astronomy 
Observatories\altaffilmark{1}} \authoraddr{Casilla 603, La Serena, Chile  \\}
\altaffiltext{1}{
Operated by the Association of Universities for Research in Astronomy, Inc.,
under cooperative agreement with the National Science Foundation.}
\begin{abstract}
We present a deep color magnitude diagram for the galactic globular cluster
NGC 2808 (C0911-646), reaching 3 magnitudes below the main sequence turnoff.
A set of local photometric standards are provided, and the mean cluster
reddening and metallicity are constrained using several photometric
techniques. The unusual horizontal branch morphology discovered by Sosin et
al. (1997, ApJ, 480, L35) is confirmed. No significant radial population
gradients are identified to within the limits of the data. We investigate
the redwards color spread on the main sequence found by Ferraro et al.
(1997, A\&A, 327, 598) in a field centered 6.3 arcmin from the cluster
center, attributed to binaries. We conclude that the spread is in the most
part due to a non-Gaussian distribution of photometric errors about the main
sequence ridgeline.
\end{abstract}

\section{INTRODUCTION}

NGC 2808 (C0911-646) is a prominent southern galactic globular cluster
situated at low galactic latitude $(l=282.^{\circ }19,b=-11.^{\circ }25).$ \
It was first noticed by Harris (1974, 1975) that the horizontal branch (HB)
morphology is very unusual, with the red HB (RHB) stars clearly separate
from a group of very blue HB (BHB) stars. In a follow-up study (Harris 1978)
the photometry was placed on a more secure basis and he suggested that the
most plausible explanation for the two groups of stars was one whereby mass
segregation of some $0.05-0.1M_{\odot }$ had occurred. His observations were
confirmed in subsequent color magnitude diagram (CMD) studies by Ferraro et
al. (1990) and by Byun \& Lee (1993). \ Gratton \& Ortolani (1986) and
Buonanno, Corsi, \& Fusi Pecci (1989) extended the CMD to below the main
sequence (MS) turn-off. \ A multi-color study by Alcaino et al. (1990)
(hereafter A90) also reached below the MS but did not include HB stars; all
these deep studies used small-format CCDs and single fields thus few evolved
stars were included in their CMDs. \ Clement \& Hazen (1989) found only two
RR Lyrae variables despite the evident richness of the total HB population.
Sosin et al. (1997) (hereafter S97) used the WFPC2 on the Hubble Space
Telescope to prepare a deep CMD for the center of NGC 2808 in the $B$ and $V$
bands. They also observed with an ultraviolet (F218W) filter, providing for
the first time good color and hence temperature resolution for the BHB
stars. The BHB was found to extend to far fainter magnitudes in the $V,B-V$
CMD than was previously known, well below the level of the MS turn-off. \
This long blue tail has two extra gaps, which are narrow and well-defined,
their width corresponding to $\sim 0.01M_{\odot }$ from comparison with HB
evolutionary models (Dorman, Rood \& O'Connell 1993). They discuss possible
explanations for the HB multi-modality, which include mass loss, composition
variations, and dynamical effects, but conclude that none of these
explanations are completely satisfactory. \ This discussion has been
extended \ by Ferraro et al. (1998)\ to other clusters that also have
extended blue HB\ . \ Finally, Ferraro et al. (1997) (hereafter F97) present
a $V,V-I$ CMD of a 4x4 arcmin region near NGC 2808, which reaches to $V\sim
23.5$ and demonstrate that the MS has dispersion greater on the redward side
of the ridge-line than on the blue side. This is interpreted as being due to
a large fraction ($\sim 24\%$) of binaries. \ 

The CMD morphology exhibited by clusters such as NGC 2808 challenges our
understanding of the advanced stages of the evolution of low-mass stars, and
how much the dense stellar environment of globular clusters affects these
processes. \ \ As the extreme example, NGC 2808 is especially important. \
The present observations complement those made with HST WFPC2, by covering
the whole cluster field at a higher photometric accuracy than that achieved
by earlier ground-based studies. The faintest magnitude in the less crowded
regions of the cluster is deeper than the HST photometry, however\
interpretation of the CMD here is complicated by the crowding, by the
significant contamination due to field stars, and by variable extinction. \ 

\section{OBSERVATIONS}

A series of CCD images centered on NGC 2808 was taken on the night of 1996
December 12, using the 4-m Blanco telescope at prime-focus. They consisted
of $V$ band (6 x 5s, 5 x 50s, 5 x 300s) and $B$ band (4 x 10s, 4 x 60s, 4 x
400s) integrations using SITe 2048 CCD \#4, and filters from set \#1. The
CCD was read out through all four amplifiers simultaneously with an Arcon
controller which delivered a read time of 35s, thus observing efficiency was
over 75\% despite the many short exposures. The 6-element prime focus
corrector, which includes prisms providing correction for atmospheric
dispersion, produces excellent quality images over an area several times
greater than the 15 arcmin square field provided by the CCD. The pixel size
of 24 microns corresponds to 0.432 arcsec and produces rather poorly sampled
images even in the worse than median seeing of this night (image fwhm 1.1
arcsec). Additionally, the CCD is not flat, however to first order the
curvature was compensated by use of a weak negative lens as dewar window.
The particular lens installed slightly under-corrected the curvature,
causing a slow degradation in focus as a function of field angle. The night
was photometric.

NGC 2808 was also observed on 1996 December 17 with the CTIO 0.9-m telescope
and Tektronix 2048 \#3 CCD, through 3 x 3 inch $B$ and $V$ filters (set Tek
2). This CCD was also read out through all four amplifiers with an Arcon
controller. Three sets of exposures were taken with differing integration
times (30s, 100s and 300s in $V$; 50s, 150s and 450s in $B$), these frames
were used for checking the photometric zeropoints, as described in detail
below. Seeing was 1.0-1.2 arcsec and the night was photometric. \ Pixel size
at the 0.9-m telescope is 0.40 arcsec.

The frames were processed in the usual way, with overscan subtraction and
trimming, followed by subtraction of a zero level exposure and removal of
the instrument signature by division using twilight sky flat-field
exposures. The zero and flat calibration frames were each made up from five
individual exposures, combined so as to remove stars from the flats and
cosmic rays from both flats and zeroes. \ Tests at the telescope using the
illuminated dome ``white-spot'' were used to measure the CCD linearity and
determine the shutter characteristics. The CCDs were measured to be linear
to better than 0.5\% (laboratory tests demonstrate linearity at the 0.1\%
level) over their operating range. Differential non-linearity between
different CCD amplifiers is much easier to measure, and is at the 0.1\%
level or better for both the CCDs used here. \ The shutters produce offsets
of +65 ms (0.9-m) and +50 ms (4-m) at their center which must be added to
the nominal exposure time, and a gradient such that by the corners of the
CCDs the offset is near zero. A mean offset was added to all exposure times,
but since all exposures were longer than 5 seconds (for which the error
induced amounts to $\pm 0.5\%$) the gradient was disregarded.

\section{DATA\ REDUCTIONS}

The individual exposures from the 4-m data-set were moved onto common
centers and combined to produce short, medium and long exposures in each
color. Integration times for the short exposures did not saturate stars at
the tip of the cluster red giant branch (RGB) and the resolution is
sufficient that bright RGB stars can be measured to the cluster center.
However this is a dense cluster, and on the longer exposures the cluster
center is hopelessly saturated, for these the central regions were blobbed
out with all pixels within a radius of 150 arcsec of the cluster center
being set to invalid. \ Reductions then proceeded using DAOPHOT II and
ALLSTAR (Stetson 1987, 1995), with the exception that a stand-alone program
was written to more critically identify PSF stars than the PICK routine in
DAOPHOT. Due to the slight variation in image quality as a function of
radius, the PSF used was a linearly varying Moffat function. Between 25000
and 40000 stars were measured on each frame, following two passes through
the FIND algorithm. Subtracted frames were clean, in particular the bright
star residuals did not differ appreciably with position, demonstrating that
the linearly varying PSF was correctly accounting for the spatial variations
in image shape. \ Stars were carefully matched between $V$ and $B$ frames,
which included accounting for a slight (approximately 0.05\%) difference in
scale.

The 0.9-m reductions were carried out in almost identical manner to those
for the 4-m. The Moffat-function PSF chosen was allowed to vary
quadratically as a function of radius, since this telescope is a classical
Cassegrain and coma is obtrusive by the corners of the CCD. \ Stars further
than 1000 pixels from the CCD center were not measured because of this
degradation in image quality.

\section{CALIBRATIONS}

The 4-m observations used the standard 4-inch square PFCCD $B$ and $V$
filters, for which color terms are accurately known. These were able to be
checked, and extinction coefficients measured, by observation of the Landolt
(1992) fields SA 98 and PG 1047 prior to the NGC 2808 observations, and the
PG 0918 field subsequently. The PG 1047 field was at airmass two, the others
at airmass similar to the cluster. Each field contains several stars with
very accurate photometry, and a total of 30 stars were measured here, using
star aperture and sky annuli radii of 14, 18 and 28 pixels respectively. The
fitted color equations are\newline
$V=v+0.019(b-v)$\newline
$B-V=0.901(b-v)$\newline
with extinction coefficients $K_{V}=0.13$ and $K_{B}=0.26$. Standard errors
of the $V$ and $B-V$ zeropoints are 0.002 and 0.003 mag. respectively.

Considering now the 0.9-m data-set, a total of 75 standards from Landolt
(1992) were measured in these two filters, and also with I filter number 31.
The standards were measured using digital aperture photometry, with star
aperture and sky annuli of 14, 18 and 28 pixels respectively. The fitted
color equations are\newline
$V=v+0.014(b-v)$\newline
$B-V=0.904(b-v)$\newline
$V-I=1.008(v-i)$\newline
with extinction coefficients $K_{V}=0.111$, $K_{B}=0.215$, and $K_{I}=0.044$%
. Standard errors of the zeropoints are 0.002 mag. in each of $V,B-V$ and $%
V-I$. A discussion of the same telescope-filter-CCD combination is given by
Walker (1995 ), from more extensive standard star observations than here. A
conclusion from that study is that errors arising from ignoring small
non-linearities in these color equations are negligible for $V$ and $V-I$,
while for $B-V$ the non-linearities amount to $\pm 0.01$ mag or less for $%
-0.3<B-V<1.8$. It was further supposed from the similarity of the 4-m and
0.9-m CCDs and filter sets that a similar conclusion could be made for the
4-m data. There is no evidence from the present data sets that color
equation non-linearities exceed these very small levels, and as such they
will be ignored. \ A caveat is that there are no Landolt standards observed
here with colors in the range $-0.2<B-V<0.15$ and we have to assume that the
color equation is well-behaved in this region. \ This is unfortunately
precisely the color range of the BHB\ stars. \ With this assumption, we
conclude that the zeropoints and color equations determined for observations
on these two nights should allow transformation to the standard system to an
accuracy of $\sim 0.01$ mag in both $V$ and $B-V$.

\subsection{NGC 2808 Local Standards}

Local standards were chosen on the 0.9m frames, and on the short-exposure
4-m frames, using the criteria of having no significant contaminating stars
within 8 arcsec radius, and be bright but not saturated. Most of these stars
are of necessity several arcmin distant from the cluster center. However we
also included stars that had been measured by A90 which are rather closer to
the cluster center. This included six stars with photoelectric photometry
from Alcaino \& Liller (1986). On the short 4-m frames two of these stars
are saturated, for the remaining four we find that in the mean, in the sense
Walker-Alcaino\newline
$\Delta V=0.01\pm 0.02$\newline
$\Delta (B-V)=0.01\pm 0.01$\newline
Similar results are found from the 0.9-m data, \newline
$\Delta V=-0.01\pm 0.02$ (n=6 short), $-0.01\pm 0.03$ (n=4 med), $-0.01\pm
0.04$ (n=3 long)\newline
$\Delta (B-V)=-0.01\pm 0.01$ (n=6), $-0.01\pm 0.01$ (n=4), $0.00\pm 0.02$
(n=3).\newline

These results are very satisfactory, however there is a strong caveat. High
contrast displays of the NGC 2808 frames clearly show that there are many
faint stars within the r=14 pixel star aperture for the stars as close to
the cluster center as the A90 standards. Thus despite the good agreement for
these stars, aperture photometry through relatively large apertures may well
be systematically incorrect. \ The new local standards set up here (Table 1)
are all more distant from the cluster and are consequently much less crowded
than the A90 stars. Tests were made by subtracting the psf-fitted faint
stars adjacent to the local standards, then comparing photometry through a
range of apertures for these frames and the originals. This showed that the
new local standards have no discernible mean zero-point offset ($\ll 0.01$
mag) in $V$ and $B-V$ when photometry with and without faint companions are
compared. However offsets for the A90 stars are typically a few 0.01 mag
when this test is performed.

An extensive set of photoelectric standards were provided by Harris (1978)
in his major study of NGC 2808. \ Most of these stars are too distant to be
on the CCD frames here. Nine stars, with $V$ magnitudes between 13.5 and
17.5 do overlap. Excluding one very discrepant star, the mean zeropoint
difference between the psf photometry here and Harris' photoelectric
photometry is, in the sense Walker-Harris,\newline
$\Delta V=-0.01\pm 0.02$ (se), (n=8) \newline
$\Delta (B-V)=0.01\pm 0.02$ (se). \newline

We made two further aperture photometry comparisons. Firstly, we compared
photometry for approximately 20 stars in the magnitude range $V=15-18$ mag
that were measured with good signal to noise on all the 0.9-m frames. No
systematic differences are found between the results from the short, medium
and long exposures, and indeed most individual comparisons agree to within
0.01-0.02 mag per star. Secondly, we also compared aperture photometry for
the same 20 stars between the 4-m short exposures and the 0.9-m long
exposures. \ The mean differences are $\ 0.00\pm 0.01$ in both $V$ and $B-V$%
. \ We calculate the formal error by adding in quadrature the mean errors
for the calibrations (linearity, flat-fielding, shutter), the errors in
color equations, zeropoints and extinction, and finally the errors in the
system of local standards. The latter group of errors are always difficult
to estimate and are usually dominated by systematic effects. Formally the
errors in $V$ and $B-V$ are little more than 0.01 mag, but we will double
these and assume that the NGC 2808 photometric system is accurate to $\pm
0.02$ mag\ in each of $V$ and $B-V$ for the purposes of the following
discussion. \ The short, medium and long exposure frames were tied together
via photometry of many (typically several hundred) overlap stars.

\subsection{\protect\bigskip Comparison of HB\ Photometry, and Variable Stars%
}

\ At this stage we checked all available blue HB photometry of NGC 2808,
since (see below) accurate photometry of the HB components is critical when
photometrically determining the cluster reddening. \ In particular, we
looked at the position of the RHB clump, and the color difference between
the RHB clump and the BHB at $V=V(HB)+1.0$ to test whether the color
differences between the various studies considered of a zeropoint shift
only, or whether they also differ in scale. \ This comparison is listed in
Table 2. \ Apparently only the Byun \& Lee (1993) photometry, which has
never been published in detail, shows both scale and zeropoint differences
when compared to the present results. The other studies show only zeropoint
shifts to within the errors. Unfortunately these zeropoint shifts are rather
large. \ Applying shifts to the Ferraro et al. (1990) photometry of $\Delta
V=+0.11,\Delta (B-V)=-0.05$ mags produces an excellent match for the whole
CMD, including the MS, with the exception that the brightest RGB stars are
0.02-0.03 mag redder here. \ For the S97 photometry, shifts of $\Delta
V=0,\Delta (B-V)=+0.05$ mag are required. \ In this case, after these shifts
are applied agreement between the two CMD's is excellent everywhere,
including the brightest RGB stars. \ 

The only search for variable stars in NGC 2808 is that by Clement \& Hazen
(1989), who from B-band photographic photometry identified two RR\ Lyrae
variables (V12 with period 0.30577705 days and\ V6 with period 0.5389687
days), a Cepheid (V10 with period 1.76528 days) and two longer period
variables (V1, V11). \ The latter two stars are confirmed to be very red,
with photometry here of \ $V=13.10,B-V=1.85$ and $V=15.11,B-V=1.27$
respectively. \ \ V1 is at the tip of the giant branch, whereas V11 is well
down the RGB. \ For V11, identification is not secure from the small-scale
Clement \& Hazen (1989) chart, and there is a very red star with $%
V=14.64,B-V=1.74$ only 3 arcsec N which may be the true variable. \ This
latter star lies well to the red of the RGB and is probably a field star. \
There are a few stars which by virtue of their position in the CMD are RR\
Lyrae candidates, but they are not numerous by comparison to the total HB
population.

\section{\protect\bigskip METALLICITY\ AND REDDENING}

Ferraro et al. (1990) provide a list of metallicity and reddening estimates
for NGC 2808. The range of $[Fe/H]$ is from -0.96 to -1.48, and of $E(B-V)$
from 0.21 to 0.34. These are not all strictly comparable, and the widely
used compilation by Zinn (1985) lists $[Fe/H]=-1.37\pm 0.09$ and $%
E(B-V)=0.22 $. Rutledge et al. (1997a,b) measured the Calcium triplet
feature for many galactic globular clusters; for NGC 2808 their mean value
is based on 20 stars and on the Zinn scale is $[Fe/H]=-1.36\pm 0.05$, in
excellent agreement with Zinn (1985).

The situation for the NGC\ 2808 reddening is not very satisfactory, with a
wide range of estimates in the recent literature. \ A90, via Vandenberg and
Bell (1985) isochrones, found $E(B-V)=0.28$ at an adopted $[Fe/H]=-1.27$,
while S97, using the Holtzman et al. (1995) filter zeropoints and color
equations, find that they can only simultaneous fit their UV and visual
CMD's with relatively small values for the reddening, between $E(B-V)=0.09$
and 0.16. \ Schlegel, Finkbeiner \& Davis (1998) find $E(B-V)=0.23$, while
Burstein \& Heiles (1984) give $E(B-V)=0.17$. The Schlegel et al. (1998)
values, plotted for a radius of 1 deg. around NGC 2208 show rather variable
reddening, ranging from $E(B-V)=0.17$ to 0.29, also note that their
zeropoint calibration produces reddenings systematically 0.02 mag greater
than those of Burstein \& Heiles (1984).

We will proceed by comparing the colors of the HB features and the RGB with
those for other clusters. \ The position of the latter is determined both by
the reddening and by the overall metallicity $[M/H].$ \ There are no
high-dispersion spectroscopic analyses available for NGC\ 2808, or indeed
for the comparison clusters we choose below, which would provide $[M/H].$ \
We make do with measurements based on the infrared Calcium triplet (Rutledge
et al. 1997a,b), strictly $[Ca/H]$, and earlier measurements (see Zinn \&
West 1984, Zinn 1985) denoted as $[Fe/H]$ but based on a number of
indicators. \ This is an unsatisfactory situation, as discussed for example
by Rutledge et al. (1997b), unlikely to be improved without more high
dispersion spectroscopic analyses for stars in many globular clusters. \ \
If at all possible, reddening should be determined \ by methods that have
little sensitivity to the metal abundance.

\ \ Unfortunately, the NGC 2808 instability strip boundaries are not well
defined thus precluding the use of the color of the blue edge which Walker
(1998) confirms to have constant $B-V$ over a wide range of metallicity, nor
are there accurate light curves and colors for the RR Lyraes which would
allow the use of Sturch's method (Sturch 1966, Walker 1990). \ Instead we
are forced mostly to metallicity-sensitive methods. \ Sarajedini \& Layden
(1997) extend their method for simultaneously determining metallicity and
reddening from a $(V,V-I)$ CMD to the $(V,B-V)$ case. Using several galactic
globular clusters with well-determined values for these two quantities as
standards they provide a calibration based on the color of the RGB at the
level of the HB\footnote{%
The level of the HB used is the mean magnitudes of the RR\ Lyraes if that is
known, else the estimated mean magnitude of the HB at the color of the
instability strip. \ For clusters with red HB morphology, the mean magnitude
of the RHB\ clump is utilised (A. Sarajedini, private communication).}, and
the magnitude difference between the RGB and the HB at $(B-V)_{0}=1.1$ and
1.2. \ The fiducial cluster NGC\ 1851 with HB level measured at the RR\
Lyrae position, and 47 Tucanae with HB\ measured from the red HB clump are
the two calibrating clusters expected to bracket NGC 2808 in metallicity, so
we will for comparison purposes measure the HB level both from the RHB\
stars and at the position of the RR\ Lyraes. \ Since NGC\ 2808 has very few
stars in the vicinity of the instability strip, there is difficulty in
estimating $V_{HB}$. \ We will use our observations of NGC\ 1851 (Walker
1992, 1998) and NGC\ 6362 (Brocato et al. 1999, Walker 1999), both of which
have $<V_{RR}>-V_{RHB}=-0.08\pm 0.01$, where for NGC\ 6362 we use the bluer
RHB clump stars only, due to the increase in $V$ magnitude range for the
redder RHB\ clump stars.\ Thus for NGC\ 2808 $<V_{RR}>=16.22$ and $%
<V_{RHB}>=16.30.$ \ \ The vertical extent of the RHB clump is 0.2 mag, so
the ZAHB level for this feature is $V=16.40\pm 0.02$. \ Although these
various HB levels are all very similar, the slope of the NGC 2808 RGB at the
HB level is approximately $\Delta B-V)/\Delta V=-0.15$ and 2.5 mags further
up the RGB\ $\Delta (B-V)/\Delta V=-0.45.$ \ Consequently, small errors in
the placement of the HB have little influence on $(B-V)_{0,g}$ but will
strongly affect any index involving the upper \ RGB. An example is the RGB
shape calibration provided by Mighell, Sarajedini \& French (1998), who
derive expressions for the color difference between the HB\ and positions 2
mags $(S_{-2.0})$ and 2.5 mags $(S_{-2.5})$ brighter, as a function of
[Fe/H]. \ 

After iterating to find consistent values from the Sarajedini \& Layden
(1997) \ fitting equations, we find $E(B-V)=0.16\pm 0.02$ and $%
[Fe/H]=-1.05\pm 0.1$ (internal errors), using $<V_{RR}>$ to represent the HB
level, and $E(B-V)=0.18\pm 0.02$ and $[Fe/H]=-1.15\pm 0.1$ (internal
errors), using $V_{RHB}$ to represent $V_{HB}.$ \ Forcing $[Fe/H]$ to a more
metal poor value consistent with the spectroscopy would increase the
reddening by $\sim 0.03$ mag, very close to the $E(B-V)=0.20\pm 0.02$
obtained from the Zinn \& West (1984) $(B-V)_{0,g}$ calibration. \ Since the
method uses a very similar $(B-V)_{0,g}$ relation to that of Zinn \& West
(1984) this result is likely telling us that the the determination of the
shape of the RGB is responsible for forcing this more metal-rich result. \ \
A \ more metal-rich value is also found from $S_{-2.0}$, $[Fe/H]=-1.10$ and $%
-1.17$ using $V_{HB}=16.22$ and 16.30 respectively. \ \ This may be
suggesting that the overall metal abundance $[M/H]$ is higher than indicated
by $[Fe/H],$ assuming $[Ca/Fe]=0,$ since it is $[M/H]$ rather than $[Fe/H]$
that determines the location and shape of the RGB. \ \ 

We can also compare the CMD to that of NGC 1851 (Walker 1992, 1998), which
has  $E(B-V)=0.02\pm 0.02$, and $[Fe/H]=-1.29\pm 0.07$, \ the latter very
similar to that for NGC 2808. \ The single Ca triplet measurement from an
integrated spectrum of the cluster (Armandroff \& Zinn 1988) implies $%
[Fe/H]=-1.16\pm 0.07$ (Da Costa \& Armandroff 1995) which is more consistent
with comparisons between positions of RGB's for various clusters. \ Despite
this uncertainty we will continue with the comparison, since NGC\ 1851, with
bimodal HB, shares the characteristic of unusual HB\ structure with NGC
2808.\ \ \ NGC\ 1851 has a very tight clump of RHB\ stars centered at $%
V=16.15,B-V=0.66,$ whereas the NGC\ 2808 RHB clump is more diffuse, and is
centered near $V=16.30,B-V=0.83.$ Forcing these to match shows that the NGC
2808 stars are $0.17\pm 0.03$ redder than the NGC 1851 stars, therefore $%
E(B-V)=0.19\pm 0.04$ . \ Comparing the two \ RGB's gives $E(B-V)=0.19\pm 0.03
$ with no correction applied for the differential difference.

In principle the BHB stars can also be overlaid, which has the advantage
that intrinsic color differences due to possibly different metallicity will
be minimized. However there are some cautions. For these very blue stars the
photometric color calibration may not be as secure as for cooler stars, as
discussed section 4. \ Additionally, any cluster-cluster variation in the
luminosity- temperature relationship for these stars means that they will
follow different loci in the CMD. \ Borissova et al. (1997) show that the
NGC 6229 and M3 sequences do not agree well, despite near-identical
reddening and metallicity, and the same is found here when comparing the NGC
1851 and NGC 2808 BHB stars. Brocato et al. (1998) give a detailed
discussion of this problem. \ More consistent results appear to be obtained
if the comparison is restricted to the color of the BHB where the sequence
begins its rapid decrease in ($V$) brightness, which occurs at $%
(B-V)=0.10\pm 0.02$ for NGC\ 1851 and at $B-V=0.30\pm 0.05$ for NGC\ 2808,
therefore $E(B-V)=0.22\pm 0.05$ for NGC 2808. \ \ As a consistency check we
generated a fiducial from the Borissova et al. (1997) CMD of NGC 6229. This
cluster has HB morphology similar to NGC 2808, with well determined and
small reddening of $E(B-V)=0.01$, and $[Fe/H]=-1.44$. Their CMD covers the
RGB and the HB only, so no comparison is possible for the turn-off region.
We normalized the $V$ magnitudes at the RGB, and shifted the NGC 6229 color
zeropoint by 0.19 mag to the red, i.e. we have assumed $E(B-V)=0.20$. We
find that the RGB of NGC 6229 is approximately 0.02 mag bluer than that of
NGC 2808 at the level of the HB, consistent with the nominal difference of
approximately 0.1 dex in the metallicities. The slope of the NGC 6229 RGB is
steeper, also consistent this cluster being somewhat more metal poor than
NGC 2808.

We can also compare with the CMD for NGC\ 6362 (Brocato et al. 1999), which
has $[Fe/H]=-1.18\pm 0.06$ (Rutledge et al. 1997b) and with slightly more
metal rich (0.1 dex) earlier estimates, all on the Zinn scale. \ Reddening,
by a variety of methods, is $E(B-V)=0.05\pm 0.02.$ \ NGC\ 6362 is one of the
most metal-rich clusters with halo kinematics, and has a well-populated HB\
with many BHB stars, but without the extensive tail of hot BHB\ stars
displayed by NGC 2808. \ With offsets applied to NGC\ 6362 fiducials of $%
\Delta V=0.92$ and $\Delta (B-V)=0.15$ mag the fit to the HB is excellent,
however the NGC\ 6362 RGB\ lies 0.04 mag redder than the NGC 2808 RGB. \
This is consistent with NGC\ 2808 being 0.2 dex more metal poor than NGC
6362.

In summary, there seems little doubt that the NGC\ 2808 \ reddening is near $%
E(B-V)=0.20\pm 0.02.$ The metal abundance, based on Calcium triplet
measurements (Rutledge at al. 1997b) is $[Fe/H]=-1.36\pm 0.05$, on the Zinn
scale. \ The shape of the RGB\ is only marginally consistent with these
values, suggesting an overall more metal-rich $[M/H]$ and a slightly smaller
value for the reddening. \ There is difficulty matching both the HB\ and the
RGB-MS in a comparison with NGC 1851, but the inconsistency is greatly
reduced if the calcium triplet-based metallicity $[Fe/H]=-1.16$ is adopted
for NGC\ 1851.

\section{COLOR\ MAGNITUDE\ DIAGRAM\ MORPHOLOGY}

The CMD morphology has been well-illustrated in the several previous studies
of this cluster, as discussed in the introduction, and we have made
comparisons with other clusters above. \ Of most interest is the HST study
by S97 which for the first time clearly shows the presence of multiple
groups of stars on the HB, and a prominent blue straggler sequence. However
even with HST the crowding in the small field near the cluster center is
sufficiently great that this CMD only reaches to $V=21$. We therefore begin
by preparing a composite CMD from the three sets of exposures. It should be
stressed that this diagram (Figure 2), containing 33464 stars, is primarily
meant to illustrate the morphology of the various sequences and does not
represent in any way the relative numbers of stars in each, since the short
exposures reached to the cluster center for the brightest stars whereas for
the longer exposures the central regions of the cluster were not measured.
It also became apparent that there were significant reddening variations
across the field, as the position of the color of the MS turn-off varied
when the data were plotted in spatial (eg 512x512 pixels) sections. As an
aside, in principle it would have been preferable to use the centroid of the
clump of RHB stars to ascertain reddening, but there were insufficient such
stars in the outer parts of the frame to provide an accurate centroid.
Photometry in the outer parts of the frame were adjusted to match the mean
of the center 1024x1024 pixels, the absolute values of the adjustments
averaged 0.028 mag for $B-V$ and were all less than 0.055 mag. Variations in
reddening across the field are not unexpected given the low galactic
latitude and the evidence for clumpy reddening in the region from the
Schlegel et al. (1998) values. The latter are consistent with the present
results, showing slightly higher reddening to the NE of NGC 2808 and
slightly lower reddening in the other three orthogonal directions.

The resultant CMD, which has been corrected for differential reddening,
clearly shows the principal sequences, contaminated by a very significant
number of field stars. When preparing this diagram it became clear that
crowding of this highly condensed cluster is the major factor in preventing
accurate ground- based photometry. Even so, the main sequence is well
defined and here extends to 4 magnitudes fainter than the turn-off. The HB
morphology is, of course, highly unusual, and we confirm the gaps discovered
by S97. Since their HB terminated near $V=21$ $B-V=-0.1$ which was their
magnitude limit, it is important to see whether or not this really is the
end of the HB. The magnitude limit of the present data is $B=25$ and $V=24$
and the HB end is seen near $V=21.2$, thus this $is$ indeed the end. There
is a  range in color for the BHB\ stars between magnitudes 16-18 that is
likely due to star-star differential reddening, also responsible for
broadening other vertical sequences with small photometric errors, such as
the RGB. \ The most prominent gap in the HB is the region of the RR Lyraes.
The RHB clump is well defined and is separated by a luminosity gap from
stars at the base of the AGB. The subgiant branch and RGB appear normal,
with the RGB terminating near $B-V=1.9$.\ 

The RGB\ bump, arising from a hiatus in the rate of stellar evolution up the
RGB as a consquence of the hydrogen burning shell passing through a
discontinuity left by the maximum penetration of the convective envelope
(Iben 1968), is clearly visible in the CMD. \ Cassisi \& Salaris (1997)
present a critical theoretical discussion of the positions of the ZAHB and
the RGB\ bump, deriving $\Delta
V_{HB}^{bump}=1.083+1.380[M/H]+0.231[M/H]^{2} $, where $[M/H]$ is the global
heavy metal abundance, which can be approximated as (Salaris, Chieffi \&
Straniero 1993) as $[M/H]=[Fe/H]+\log (0.638\cdot f+0.362)$, where $\log
f=[\alpha /Fe]$, the enhancement factor for the $\alpha $ elements. \ They
also find that the position of the bump depends on the mass of the model,
and derive an increase of $\Delta V_{HB}^{bump}$ of almost 0.024 mag for an
increase of 1 Gyr in age of the cluster. \ In an observational evaluation,
Saviane et al. (1998) \ approximate the relation as \ $\Delta
V_{HB}^{bump}=0.66[Fe/H]+0.87.$ \ The HB reference magnitude in both cases
is that for the ZAHB in the vicinity of the RR\ Lyraes. \ We use the
comparison \ with NGC\ 1851 and 6362 (see above) to estimate $V_{ZAHB}$ for
NGC\ 2808, thus obtaining $\Delta V_{HB}^{bump}=0.00\pm 0.07$, and with the
S98 calibration $[Fe/H]=-1.32\pm 0.11$. \ With the Cassisi \& Salaris (1997)
formulation, we find $[M/H]=-0.93\pm 0.07.$ \ Since there is no measured
value of $[\alpha /Fe]$ for NGC 2808, we will adopt a value of $0.30\pm 0.05$%
, whereupon $[Fe/H]=-1.14\pm 0.08.$ \ As pointed out by Cassisi \& Salaris
(1997), their theoretical scale agrees well with high-dispersion
spectroscopic measurements of cluster metallicities, which tend to give
values somewhat more metal-rich than the Zinn scale, the Zinn-scale estimate
of $[Fe/H]=-1.36\pm 0.05$ transforms to $[Fe/H]=-1.11\pm 0.03$ on the
Carretta \& Gratton (1997) high dispersion spectroscopic scale (Rutledge et
al. 1997b). \ So in this admittably round-about fashion, the position of the
NGC\ 2808 RGB\ bump is found to be in excellent agreement with the Cassisi
\& Salaris (1997) models, assuming the $[\alpha /Fe]$ ratio is normal. \
This would also argue against enhanced $[\alpha /Fe]$ being responsible for
the shallower than expected RGB, as discussed above.

\section{RADIAL POPULATION GRADIENTS\qquad}

The measurement of the relative numbers of stars as a function of radius in
a cluster as rich as NGC 2808 is a difficult task, and one best suited to
HST. Here we have made a few comparisons between groups of stars of roughly
equivalent brightness in order to minimize magnitude-dependent completeness
corrections. The extent of the difficulty with the present data set is
evident by comparing the ratio of AGB+RGB stars brighter than $V=15$ with
the RGB stars with $V=15$ to 17.5, using only the short exposure frames.
Outside $r=150$ pixels\ (one pixel = 0.432 arcsec) the ratio faint:bright is
constant at 8.8, whereas from 50:150 pixels it is 5.4, and inside 50 pixels
it is 1.5. We find no significant differences between the distribution of
RHB and BHB stars within the errors, comparing only the brightest clump ($%
V<18.5$) of BHB stars; there are $2.1\pm 0.5$ times as many RHB stars as
these BHB stars in the $150-300$ pixel radii annulus and this ratio is
preserved, within a progressively larger error, at greater distances. For
the two fainter clumps of BHB stars the number counts would appear to
indicate a deficiency of such stars at small radii, compared to the RHB or
brighter BHB stars, but a comparison between ratios of MS stars with the
same $V$ magnitude shows that the number counts of stars as faint as this
are seriously incomplete in the inner annuli. \ We do find that the HB star
distribution in the outer annuli (ie at distances greater than 300 pixels
from the cluster center, is similar to that seen at the cluster center by
S97. This would argue for there being no gross mass differences between the
sub-groups on the HB, or else the cluster is dynamically mixed. The latter
could be tested by comparing the radial distribution of the BS stars to that
for other cluster stars.

\section{BINARIES\qquad}

The F97 field is about 6 arcmin south and 2 arcmin east of the cluster
center, and their CMD shows a redwards broadening of the MS that can be
interpreted as being due to binaries, with some $24\pm 4\%$ of apparently
single MS stars being in such a category. As cautioned by F97 these may be
unresolved optical binaries, some of which must be expected due to the high
star density, rather than true physical binaries. Unfortunately it is
difficult to directly test between the two options, as both mass segregation
of the binary systems (e.g. Rubenstein \& Bailyn 1999) and the effects of
image blending will act to concentrate such stars towards the center of the
cluster. F97 consider, from artificial star tests, that no more than 10\% of
the candidate binaries are optical in character.

We first compare the two extremes. Stars nearer than 350 pixels (2.52
arcmin) from the cluster center (note that no stars from the long exposures
were measured this close) are plotted in Figure 3. Of the 3566 stars in this
CMD, only 300 are on the MS ($V>20$) and these clearly suffer much
photometric scatter. However there is very little field star contamination
and the evolved star sequences, with the exception of the fainter BHB stars,
are well populated. There are at least 25 BS stars. But from this CMD
nothing can be said about the spread of MS stars. We can contrast this
(Figure 4) with a CMD made up by selecting stars more than 1000 pixels from
the cluster center. This shows a well defined cluster MS, few evolved stars,
and the expected large number of field stars. A careful examination of the
distribution of stars about a MS fiducial shows no evidence for any
significant binary population. Turning now to fields at intermediate radii,
we plot the CMD between radii of $600-800$ pixels (Figure 5) and $800-1000$
pixels (Figure 6). It is visually obvious from the CMDs, especially Figure
5, that there is a redwards spread of the MS. We therefore need to carefully
test the stellar image crowding in these annuli in order to differentiate
between the optical and physical binary hypotheses. The number of measured
stars in each CMD is 9119 and 6753; thus there are 96 and 167 pixels per
star respectively. With image fwhm of under 2.5 pixels, this degree of
crowding does not seem excessive, and indeed with 1 arcsec seeing the image
crowding for these data becomes severe only well inside the $r=600$ pixels
radius. To put this initial impression on a more quantitative basis,
artificial stars were added to the long-exposure frames using the ADDSTAR
routine of DAOPHOT (Stetson 1987). As long as the brightness of the added
stars exceeded the MS magnitude limit of $V\sim 23.5$ by more than one
magnitude, the percentage of stars recovered in both annuli exceeded 95\%.
It is therefore clear, as was concluded by F97, that optical binaries make
only a minor contribution to the observed distribution. Our next test was to
restrict the analysis to only the stars in the $600-800$ pixel annulus. Note
that this annulus is slightly closer in the mean to the cluster center (5
arcmin) than the F97 field,

We have already purged the star list of stars with poor profile fit and
elliptical shape parameters, as determined by the DAOPHOT PSF fitting, and
have also cut-off stars with standard errors (as produced by ALLSTAR)
greater than twice the median standard error of all stars in the sample, in
one magnitude wide bins. For stars on the MS, the cutoff corresponds to a
standard error  $\sigma (B-V)=(0.06,0.08,0.10,0.15)$ mag. at $V=(20,21,22,23)
$ mag. respectively. \ We call this set of stars ``sample one'', and it
corresponds to what is shown in Figure 5. We then further excluded all stars
with standard errors greater than 0.030 mag, which had the effect of
removing all stars fainter than $V=22.6$ as well as the significant number
of stars with color errors between 0.03 mag and the cut-offs given above.
The remaining stars are ``sample two''. The radial distributions of the
stars in the two samples differ negligibly, with less than one pixel
difference in mean distance from the cluster center. \ We then compared the
color distributions across the same MS fiducial, from $V=20.0-22.0$,
normalizing the two distributions at the peak. The normalizing factor (1.16)
is not too different from unity, and the resulting distributions are plotted
in Figure 7. \ Bluewards of the peak the two curves are nearly identical.
The redward distributions are very different, with the{\it \ low error }plot
(sample 2) showing substantially fewer stars and a distribution close to
symmetrical across the peak, especially if the field star contributions are
taken into account. \ Given the ADDSTAR experiments discussed above, some of
small number of excess stars (the red side of the sample two distribution)
may be optical binaries, but even if all the outliers are physical binaries
they contribute no more than $10\%$ of the total number of stars measured on
the MS. The distribution is however, equally consistent with a MS binary
proportion of zero.

These experiments show that within the annulus under consideration, the
apparent binary population is due to a non-Gaussian distribution of color
errors about the MS fiducial. An explanation could be that the most likely
color of any contaminating star is redder than a given MS star, since
fainter stars are either cool cluster MS stars or very red foreground
dwarfs, or faint compact galaxies. \ We therefore conclude that there is no
clear evidence from the MS color distribution to support the existence of a
large proportion of physical binaries in NGC 2808, at least outside a radius
of a few arcmin.

\section{\protect\bigskip FINAL\ REMARKS}

NGC\ 2808 has the most unusual HB\ morphology of any galactic GC, and is an
important fiducial for any explanation of the late stages of stellar
evolution in general and in high-density GC in particular. \ In this paper,
based on photometrically secure observations, we determine the cluster
reddening to be $E(B-V)=0.20\pm 0.02$ and show that this is consistent with
the spectroscopically determined metallicity. \ Small differences in the
values for $E(B-B)$ and $[Fe/H]$ via different methods may only reflect
calibration difficulties, but may also be indicative of differences in $%
[\alpha /Fe]$ between the comparison clusters chosen to have similar $[Fe/H]$
to NGC 2808. \ Significant progress in this regard requires high dispersion
spectroscopy in order to determine total metal abundance $[M/H]$. \ \ The
CMD presented confirms and extends recent HST results, although there is an
0.05 mag difference in the $B-V$ photometric zeropoint. \ No significant
differences in radial gradients between various groups of stars on the CMD
are found, while the presence of a large fraction of MS binaries is not
confirmed. \ However the observations here do not extend into the core
region of this very dense cluster, and studies of gradients and binary
fraction to the very center of NGC 2808, using HST, would be of considerable
interest.

\newpage

\begin{figure}
\caption{NGC 2808, with the local standards identified, from a 5 second V
band exposure with the CTIO 4-m PFCCD camera.  The field size is 14.75 x 14.75 arcmin, 
with north to the right and east up.  The scale units are 4 pixels, or 1.728 arcsec.}
\end{figure}

\begin{figure}
\caption{Color-magnitude diagram for all stars in the 4-m field. The data
plotted are a combination of measurements on the short, medium and long exposures, 
using a weighted mean for cases of multiple measurement.  The photometry has been corrected
for differential reddening with respect to the center of the field.}
\end{figure}

\begin{figure}
\caption{Color-magnitude diagram for stars closer than 350 pixels (2.5 arcmin)
from the center of NGC 2808.}
\end{figure}

\begin{figure}
\caption{Color-magnitude diagram for stars more than 1000 pixels (7.2 arcmin)
distant from the center of NGC 2808.}
\end{figure}

\begin{figure}
\caption{Color-magnitude diagram for stars measured within an annulus of
radii 600 and 800 pixels (4.3 and 5.8 arcmin) centered on NGC 2808.}
\end{figure}

\begin{figure}
\caption{Color-magnitude diagram for stars measured within an annulus of
radii 800 and 1000 pixels (5.8 and 7.2 arcmin) centered on NGC 2808.}
\end{figure}

\begin{figure}
\caption{Normalized histograms showing the distribution of stars around a
fiducial centered on the MS, from $V = 20-22$.  See text for details.}
\end{figure}

\clearpage

TABLE CAPTIONS

Table 1:   NGC 2808 Local Standards

Table 2:   HB Color Location

\begin{deluxetable}{lllll}
\footnotesize
\scriptsize
\tablewidth{0pt}
\tablenum{1}
\tablecaption{NGC 2808 Local Standards}
\tablehead{
\colhead{Num}  &
\colhead{X}   &
\colhead{Y}   &
\colhead{$V$} &
\colhead{$B-V$} }
\startdata
  1 & 1043.94 &  548.13   &    13.544  & 0.942 \nl
  2 & 1234.07 &  276.91   &    14.425  & 1.136 \nl
  3 & 1020.65 &  507.75   &    15.091  & 1.134 \nl
  4 & 1416.60 &  197.26   &    15.848  & 0.915 \nl
  5 &  189.81 & 1749.83   &    13.444  & 1.930 \nl
  6 &  167.40 & 1879.44   &    14.747  & 0.436 \nl
  7 &  451.61 & 1712.06   &    13.933  & 1.077 \nl
  9 &  282.54 & 1545.45   &    13.085  & 0.569 \nl
 10 &   58.69 &  967.08   &    14.919  & 0.586 \nl
 11 &  218.19 &  514.51   &    15.737  & 0.882 \nl
 12 &  457.55 &  264.05   &    14.591  & 0.696 \nl
 13 & 1007.26 &  122.55   &    14.417  & 1.384 \nl
 14 &  771.75 &   41.31   &    14.318  & 1.409 \nl
 15 & 1581.32 &   76.12   &    13.627  & 1.271 \nl
 16 & 1839.59 &  219.65   &    13.389  & 0.476 \nl
 18 & 1783.94 &  454.64   &    14.613  & 1.287 \nl
 19 & 1995.65 &  871.26   &    13.951  & 0.609 \nl
 21 & 1855.74 & 1684.73   &    14.423  & 0.852 \nl
 23 & 1063.22 & 1779.47   &    13.841  & 0.833 \nl
 \enddata
 \end{deluxetable}
\begin{deluxetable}{llll}
\footnotesize
\tablewidth{0pt}
\tablenum{2}
\tablecaption{HB Color Location}
\tablehead{
\colhead{Source}  &
\colhead{RHB}   &
\colhead{BHB}   &
\colhead{Difference} }
\startdata
This paper  & 0.83    &  0.13    &  0.70  \nl
Sosin et al & 0.78    &  0.08    &  0.70  \nl
Ferraro     & 0.88    &  0.17    &  0.71  \nl
Alcaino     & 0.80    &    -     &   -    \nl
Byun \& Lee & 0.82    &  0.06    &  0.76  \nl
 \enddata
 \end{deluxetable}
 \end{document}